\documentclass[tightenlines,superscriptaddress,10pt,twocolumn,aps,prd]{revtex4-1}

\usepackage{geometry}
\usepackage{graphicx,wrapfig}
\usepackage[export]{adjustbox}
\usepackage{epsfig}
\usepackage{float}
\usepackage{flafter}
\usepackage{subfigure}
\usepackage{sidecap}
\usepackage{notes2bib}
\usepackage{gensymb}
\usepackage{url}
\usepackage{amsmath}
\usepackage{latexsym}
\usepackage{amssymb}
\usepackage{xcolor}
\usepackage{color}
\usepackage{slashed}
\usepackage{units}
\usepackage{siunitx}
\usepackage{notoccite}
\usepackage[utf8]{inputenc}
\usepackage[english]{babel}
\usepackage{csquotes}
\usepackage{dcolumn}
\usepackage{bm}
\usepackage{color}
\usepackage{enumitem}
\usepackage{pifont}
\usepackage{textcomp}
\usepackage{tikz}
\usepackage{colortbl}
\usepackage[version=4]{mhchem}
\usepackage{xspace}
\usepackage{placeins}
\usepackage{tabularx}
\usepackage{hyperref}
\usepackage{amssymb}
\usepackage{hyperref}
\usepackage{lineno}
\usepackage{tikz-feynman}

\usepackage{pgfgantt}
\usepackage[ampersand]{easylist}

\begin{document}

\onecolumngrid

\title{A COHERENT constraint on leptophobic dark matter using CsI data}

\widowpenalty10000
\clubpenalty10000
\renewcommand\floatpagefraction{1}
\renewcommand\topfraction{1}
\renewcommand\bottomfraction{1}
\renewcommand\textfraction{0}

\setlength{\belowcaptionskip}{-10pt} 

\renewcommand{\thesection}{\arabic{section}}
\renewcommand{\thesubsection}{\thesection.\arabic{subsection}}
\renewcommand{\thesubsubsection}{\thesubsection.\arabic{subsubsection}}

\makeatletter
\renewcommand{\p@subsection}{}
\renewcommand{\p@subsubsection}{}
\makeatother

\newcommand{\Mephidesc}{\affiliation{National Research Nuclear University MEPhI (Moscow Engineering Physics Institute), Moscow, 115409, Russian Federation}}
\newcommand{\Dukedesc}{\affiliation{Department of Physics, Duke University, Durham, NC, 27708, USA}}
\newcommand{\TUNLdesc}{\affiliation{Triangle Universities Nuclear Laboratory, Durham, NC, 27708, USA}}
\newcommand{\UTKdesc}{\affiliation{Department of Physics and Astronomy, University of Tennessee, Knoxville, TN, 37996, USA}}
\newcommand{\ITEPdesc}{\affiliation{National Research Center  ``Kurchatov Institute'' Kurchatov Complex for Theoretical and Experimental Physics, Moscow, 117218, Russian Federation}}
\newcommand{\ORNLdesc}{\affiliation{Oak Ridge National Laboratory, Oak Ridge, TN, 37831, USA}}
\newcommand{\USDdesc}{\affiliation{Physics Department, University of South Dakota, Vermillion, SD, 57069, USA}}
\newcommand{\NCSUdesc}{\affiliation{Department of Physics, North Carolina State University, Raleigh, NC, 27695, USA}}
\newcommand{\Sandiadesc}{\affiliation{Sandia National Laboratories, Livermore, CA, 94550, USA}}
\newcommand{\UWdesc}{\affiliation{Center for Experimental Nuclear Physics and Astrophysics \& Department of Physics, University of Washington, Seattle, WA, 98195, USA}}
\newcommand{\LANLdesc}{\affiliation{Los Alamos National Laboratory, Los Alamos, NM, 87545, USA}}
\newcommand{\Laurentiandesc}{\affiliation{Department of Physics, Laurentian University, Sudbury, Ontario, P3E 2C6, Canada}}
\newcommand{\CMUdesc}{\affiliation{Department of Physics, Carnegie Mellon University, Pittsburgh, PA, 15213, USA}}
\newcommand{\IUdesc}{\affiliation{Department of Physics, Indiana University, Bloomington, IN, 47405, USA}}
\newcommand{\VTdesc}{\affiliation{Center for Neutrino Physics, Virginia Tech, Blacksburg, VA, 24061, USA}}
\newcommand{\NCCUdesc}{\affiliation{Department of Mathematics and Physics, North Carolina Central University, Durham, NC, 27707, USA}}
\newcommand{\NCSUnucengdesc}{\affiliation{Department of Nuclear Engineering, North Carolina State University, Raleigh, NC, 27695, USA}}
\newcommand{\UFdesc}{\affiliation{Department of Physics, University of Florida, Gainesville, FL, 32611, USA}}
\newcommand{\Tuftsdesc}{\affiliation{Department of Physics and Astronomy, Tufts University, Medford, MA, 02155, USA}}
\newcommand{\SNUdesc}{\affiliation{Department of Physics and Astronomy, Seoul National University, Seoul, 08826, Korea}}
\author{D.~Akimov}\Mephidesc
\author{P.~An}\Dukedesc\TUNLdesc
\author{C.~Awe}\Dukedesc\TUNLdesc
\author{P.S.~Barbeau}\Dukedesc\TUNLdesc
\author{B.~Becker}\UTKdesc
\author{V.~Belov }\ITEPdesc\Mephidesc
\author{I.~Bernardi}\UTKdesc
\author{M.A.~Blackston}\ORNLdesc
\author{C.~Bock}\USDdesc
\author{A.~Bolozdynya}\Mephidesc
\author{R.~Bouabid}\Dukedesc\TUNLdesc
\author{J.~Browning}\NCSUdesc
\author{B.~Cabrera-Palmer}\Sandiadesc
\author{D.~Chernyak}\altaffiliation{Now at:  Department of Physics and Astronomy, Tuscaloosa and Institute for Nuclear Research of NASU, Kyiv, 03028, Ukraine}\USDdesc
\author{E.~Conley}\Dukedesc
\author{J.~Daughhetee}\ORNLdesc
\author{J.~Detwiler}\UWdesc
\author{K.~Ding}\USDdesc
\author{M.R.~Durand}\UWdesc
\author{Y.~Efremenko}\UTKdesc\ORNLdesc
\author{S.R.~Elliott}\LANLdesc
\author{L.~Fabris}\ORNLdesc
\author{M.~Febbraro}\ORNLdesc
\author{A.~Gallo Rosso}\Laurentiandesc
\author{A.~Galindo-Uribarri}\ORNLdesc\UTKdesc
\author{M.P.~Green }\TUNLdesc\ORNLdesc\NCSUdesc
\author{M.R.~Heath}\ORNLdesc
\author{S.~Hedges}\Dukedesc\TUNLdesc
\author{D.~Hoang}\CMUdesc
\author{M.~Hughes}\IUdesc
\author{B.A.~Johnson}\IUdesc
\author{T.~Johnson}\Dukedesc\TUNLdesc
\author{A.~Khromov}\Mephidesc
\author{A.~Konovalov}\Mephidesc
\author{E.~Kozlova}\Mephidesc\ITEPdesc
\author{A.~Kumpan}\Mephidesc
\author{L.~Li}\Dukedesc\TUNLdesc
\author{J.M.~Link}\VTdesc
\author{J.~Liu}\USDdesc
\author{A.~Major}\Dukedesc
\author{K.~Mann}\NCSUdesc
\author{D.M.~Markoff}\NCCUdesc\TUNLdesc
\author{J.~Mastroberti}\IUdesc
\author{J.~Mattingly}\NCSUnucengdesc
\author{P.E.~Mueller}\ORNLdesc
\author{J.~Newby}\ORNLdesc
\author{D.S.~Parno}\CMUdesc
\author{S.I.~Penttila}\ORNLdesc
\author{D.~Pershey}\email{daniel.pershey@duke.edu}\Dukedesc
\author{C.~Prior}\Dukedesc\TUNLdesc
\author{R.~Rapp}\CMUdesc
\author{H.~Ray}\UFdesc
\author{O.~Razuvaeva}\Mephidesc\ITEPdesc
\author{D.~Reyna}\Sandiadesc
\author{G.C.~Rich}\TUNLdesc
\author{J.~Ross}\NCCUdesc\TUNLdesc
\author{D.~Rudik}\Mephidesc
\author{J.~Runge}\Dukedesc\TUNLdesc
\author{D.J.~Salvat}\IUdesc
\author{A.M.~Salyapongse}\CMUdesc
\author{J.~Sander}\USDdesc
\author{K.~Scholberg}\Dukedesc
\author{A.~Shakirov}\Mephidesc
\author{G.~Simakov}\Mephidesc\ITEPdesc
\author{W.M.~Snow}\IUdesc
\author{V.~Sosnovstsev}\Mephidesc
\author{B.~Suh}\IUdesc
\author{R.~Tayloe}\IUdesc
\author{K.~Tellez-Giron-Flores}\VTdesc
\author{I.~Tolstukhin}\altaffiliation{Now at: Argonne National Laboratory, Argonne, IL, 60439, USA}\IUdesc
\author{E.~Ujah}\NCCUdesc\TUNLdesc
\author{J.~Vanderwerp}\IUdesc
\author{R.L.~Varner}\ORNLdesc
\author{C.J.~Virtue}\Laurentiandesc
\author{G.~Visser}\IUdesc
\author{T.~Wongjirad}\Tuftsdesc
\author{Y.-R.~Yen}\CMUdesc
\author{J.~Yoo}\SNUdesc
\author{C.-H.~Yu}\ORNLdesc
\author{J.~Zettlemoyer}\altaffiliation{Now at: Fermi National Accelerator Laboratory, Batavia, IL, 60510, USA}\IUdesc

\begin{abstract}
We use data from the COHERENT CsI[Na] scintillation detector to constrain sub-GeV leptophobic dark matter models.  This detector was built to observe low-energy nuclear recoils from coherent elastic neutrino-nucleus scattering.  These capabilities enable searches for dark matter particles produced at the Spallation Neutron Source mediated by a vector portal particle with masses between 2 and 400~MeV/c$^2$.  No evidence for dark matter is observed and a limit on the mediator coupling to quarks is placed.  This constraint improves upon previous results by two orders of magnitude.  This newly explored parameter space probes the region where the dark matter relic abundance is explained by leptophobic dark matter when the mediator mass is roughly twice the dark matter mass.  COHERENT sets the best constraint on leptophobic dark matter at these masses.
\end{abstract}

\maketitle

\clearpage

\twocolumngrid

\section{Introduction}

There is overwhelming evidence for the gravitational effects of dark matter which comprises $\approx$~80$\%$ of the matter in the universe~\cite{Freese:2017idy}.  Despite several experimental techniques developed to detect dark matter, its particle nature is still not understood.  To resolve this question, several experimental approaches have been attempted to identify dark-matter particles using both astroparticle~\cite{PhysRevLett.121.111302,PhysRevLett.123.251801,XENON:2020fgj} and accelerator-based techniques~\cite{Aguilar-Arevalo:2017mqx,Aguilar-Arevalo:2021sbh}.  

COHERENT detectors deployed at the Spallation Neutron Source (SNS) are sensitive to dark-matter particles produced in the target with masses below the current beam energy, $\approx1$~GeV.  From cosmological constraints, sub-GeV dark matter cannot interact directly with standard-model particles through the weak force~\cite{PhysRevLett.39.165}.  Instead, light dark matter would consist of hidden sector particles, $\chi$, whose interactions with standard-model particles are mediated by a vector portal particle $V$.  Sub-GeV dark matter may be scalar or fermionic, and there are different channels through which $V$ could interact with standard-model particles.  

Detectors sensitive to low-energy nuclear recoils induced by coherent elastic neutrino nucleus scattering (CEvNS) are efficient probes of light dark matter with masses below $\approx1$~GeV.  These detectors would also observe nuclear recoils induced by coherent $\chi$-nucleus scattering if dark matter is produced at the SNS.  The cross section for this process is proportional to the square of the nucleon number, $A$~\cite{PhysRevD.80.095024} so that a small detector can yield a result competitive with constraints from much larger detectors that rely on inelastic signal channels.  Additionally, as accelerator-produced light dark matter is relativistic, the scattering cross section is relatively independent of dark matter spin~\cite{Battaglieri:2017aum} so that CEvNS detectors can effectively search for both scalar and fermionic dark matter.  

\section{COHERENT at the SNS}

The COHERENT collaboration employs several detectors at the SNS at Oak Ridge National Laboratory~\cite{COHERENT:2018gft}.  We measure CEvNS and other low-energy scattering processes on several types of nuclei and maintain neutron detectors to understand beam-related backgrounds.  Detectors are placed in a basement hallway, Neutrino Alley, where neutron backgrounds are low enough to permit these measurements of low-energy scattering processes about 20~m from the SNS target.  The SNS is a $\pi$ decay-at-rest ($\pi$-DAR) source with a FWHM beam pulse width of 360~ns offering a prompt $\nu_\mu$ flux and delayed $\nu_e/\bar{\nu}_\mu$ flux separated in time.  Having multiple flavors and separation by timing is ideal for testing beyond-the-standard-model (BSM) scenarios such as lepton flavor universality of the CEvNS cross section (at tree level) and for searches for hidden sector particles such as dark matter.

To study CEvNS, we have built several detectors sensitive to low-energy nuclear recoils.  The COHERENT CsI[Na] detector was a 14.6~kg, low-background scintillating crystal commissioned at the SNS which made the first observation of CEvNS in 2017~\cite{Akimov:2017ade}.  Scintillation light was collected by a single Hamamatsu R877-100 photomultiplier.  Neutron and $\gamma$ backgrounds were mitigated by a composite shielding using both low-activity lead and low-$Z$ materials.  The assembly was surrounded by a plastic scintillator veto to reject cosmic-induced activity.  The detector collected beam data from 2017 to 2019, recording a total of $3.20\times10^{23}$~protons-on-target during its run.  Recently, the full dataset from the detector, along with improved understanding of the detector response to nuclear recoils, enabled the most precise measurement of the CEvNS cross section yet~\cite{Akimov:2021dab} and placed leading constraints on sub-GeV dark matter~\cite{COHERENT:2021pvd}.  

Additionally, we study CEvNS on Ar with continuing operation of a liquid argon scintillation detector.  This detector made the first measurement of the CEvNS cross section on Ar~\cite{Akimov:2020pdx} with an active argon mass of 24~kg.  COHERENT is also currently commissioning low-threshold CEvNS detectors with Ge and Na targets along with a heavy-water Cherenkov detector to calibrate the neutrino flux at the SNS~\cite{Akimov_2021}.  Neutron detectors are currently running to monitor beam-related backgrounds~\cite{COHERENT:2021qbu}.  COHERENT also plans for future large-scale detectors which will dramatically improve precision of measurements on small, first-light detectors~\cite{Akimov:2022oyb}.

\section{Leptophobic dark matter}

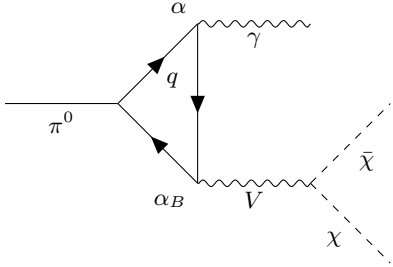
\begin{figure}[H]
\centering
\begin{tikzpicture}{}
    \begin{feynman}
        \vertex(a);
        \vertex [right=of a] (b);
        \vertex [above right=of b] (d);
        \vertex [right=of d] (e);
        \vertex [below right=of b] (f);
        \vertex [right=of f] (g);
        \vertex [above right=of g] (h);
        \vertex [below right=of g] (i);
        \vertex [above right=of i] (j);
        \vertex [left=of i] (k);
        \vertex [below right=of i] (l);
        \diagram*{
            (a) -- [plain, edge label'=\(\pi^0\)] (b),
            (b) -- [fermion, edge label'=\(q\)] (d),
            (d) -- [fermion] (f),
            (f) -- [fermion] (b),
            (d) -- [photon, edge label'=\(\gamma\)] (e),
            (f) -- [boson, edge label'=\(V\)] (g),
            (g) -- [scalar, edge label'=\(\chi\)] (i),
            (g) -- [scalar, edge label'=\(\bar{\chi}\)] (h),
        };
        \vertex [above left=0.15em of d] {\(\alpha\)};
        \vertex [below left=0.15em of f] {\(\alpha_B\)};
    \end{feynman}
\end{tikzpicture}
\caption{Production of leptophobic dark matter from $\pi^0$ decay which may occur at the SNS.  The dark matter particles are on-shell and may subsequently scatter coherently in a COHERENT detector.}
\label{fig:FeynmanPr}
\end{figure}

COHERENT has previously reported a dark-matter constraint using these data on a model that predicts a vector portal particle $V$ that kinetically mixes with the photon and decays to dark-matter particles $\chi$~\cite{COHERENT:2021pvd}.  A leptophobic, or baryonic, dark-matter model~\cite{Aranda:1998fr,Gondolo2012LightDM,Batell:2014yra,deNiverville:2015mwa,deNiverville:2016rqh,Boyarsky:2021moj} is also viable where $V$ mediates interactions between $\chi$ and quarks described in terms of the vector mediator, $V^\mu$, and the scalar dark matter particle, $\chi$, by Lagrangian terms
\begin{equation}
    \mathcal{L} \supset -g_BV^\mu J_\mu^B+g_\chi V^\mu (\partial_\mu\bar{\chi}\chi-\bar{\chi}\partial_\mu\chi) 
\end{equation}
where $J_\mu^B$ is the baryon current given by 
\begin{equation}
J_\mu^B=\frac{1}{3}\sum_{q}\bar{q}\gamma_\mu q.
\end{equation}
This implies the couplings $\alpha_B\equiv g_B^2/4\pi$ and $\alpha_\chi$~$\equiv$~$g_\chi^2/4\pi$ describing $Vqq$ and $V\chi\chi$ vertices, respectively.  Production of leptophobic dark matter can be achieved in $\pi^0$ decay by the diagram shown in Fig.~\ref{fig:FeynmanPr} with $\text{Br}(\pi^0\rightarrow\gamma V)\propto2\alpha_B/\alpha$.  Coherent $\chi-A$ scattering off a nucleus, $A$, occurs through simple $V$ exchange with $\sigma(\chi A\rightarrow\chi A)\propto \alpha_B\alpha_\chi$.

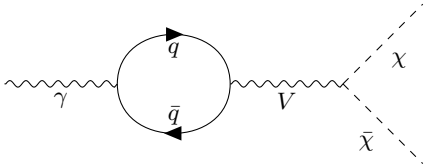
\begin{figure}[H]
\centering
\begin{tikzpicture}{}
    \begin{feynman}
        \vertex(a);
        \vertex [right=of a] (b);
        \vertex [right=of b] (c);
        \vertex [right=of c] (d);
        \vertex [above right=of d] (e);
        \vertex [below right=of d] (f);
        \diagram*{
            (a) -- [photon, edge label'=\(\gamma\)] (b),
            (b) -- [fermion, half left, edge label'=\(q\)] (c) -- [fermion, half left, edge label'=\(\bar{q}\)] (b),
            (c) -- [boson, edge label'=\(V\)] (d),
            (d) -- [scalar, edge label'=\(\chi\)] (e),
            (d) -- [scalar, edge label'=\(\bar{\chi}\)] (f),
        };
    \end{feynman}
\end{tikzpicture}
\caption{The $\gamma-V$ kinetic mixing induced from leptophobic dark-matter portal through a virtual quark loop and subsequent $V\rightarrow\bar{\chi}\chi$ decay.}
\label{fig:Feynman}
\end{figure}

Though there is no $V-\gamma$ kinetic mixing in the leptophobic dark matter model at tree level, there is an effective mixing from the loop diagram shown in Fig.~\ref{fig:Feynman} with a mixing parameter $\varepsilon\sim eg_B/(4\pi)^2$.  Through this effective kinetic mixing, the couplings for leptophobic dark matter can be related to the dark-matter scattering cross section at freeze-out which determines the modern relic abundance.  

There are several current constraints on this model from Coherent Captain Mills~\cite{Aguilar-Arevalo:2021sbh}, NA62~\cite{NA62:2019meo}, MiniBooNE~\cite{MiniBooNE:2017nqe,MiniBooNEDM:2018cxm}, and neutron-scattering~\cite{BARBIERI1975270}.  There is also a model-dependent anomalon limit considering the influences of dark matter on anomalous SM baryonic couplings~\cite{Dror:2017ehi}.  These constraints exclude leptophobic dark matter for all values of $\alpha_\chi$ if $m_V/m_\chi>3$.  

However, there is significant parameter space viable for $m_V/m_\chi\approx2$.  If $0<m_V-2m_\chi<T_f\approx m_\chi/20$, with $T_f$ the freeze-out temperature, then the annihilation rate of dark matter would have happened on resonance~\cite{PhysRevD.96.095022} in the early universe during thermal freeze-out, increasing the $\bar{\chi}\chi\rightarrow\text{SM}$ cross section.  As the annihilation cross section can be determined from the relic dark matter density, this resonance implies that model couplings required to produce the observed relic density are much lower than when off resonance, $m_V/m_\chi\geq3$.  Thus experimental searches for leptophobic dark matter in experiments must probe significantly lower couplings if $m_V/m_\chi$ is slightly above 2.  The resonant enhancement of the scattering cross section at freeze-out is parameterized in terms of $\varepsilon_R\equiv(m_V^2-4m_\chi^2)/4m_\chi^2$.  Thus, couplings required to match the relic abundance of dark matter also depend on $\varepsilon_R$ with decreasing target couplings as $\varepsilon_R\rightarrow0$.  For scalar leptophobic dark matter, this effect reaches a floor for $\varepsilon_R<10^{-5}$ where lower values of $\varepsilon_R$ do not further suppress expected dark matter couplings.  For fermionic dark matter, however, this effect is unbounded with $\varepsilon_R\rightarrow0$ driving the coupling required to match the relic abundance to arbitrarily low values.

The dominant production mechanisms for leptophobic dark matter at the SNS are $\pi^0\rightarrow\gamma V$ and $\eta^0\rightarrow\gamma V$ facilitated through a $Vqq$ vertex for the SNS beam energy as was the case with the leptophilic model~\cite{COHERENT:2021pvd}.  A leptophobic dark matter particle passing through the CsI[Na] detector can scatter coherently with target nuclei, producing a nuclear recoil signature with similar energies to those expected from CEvNS.  The differential cross section is given in terms of the nuclear mass, $m_N$, recoil energy, $E_r$, nuclear form factor, $F(Q^2)$, and momentum transfer, $Q^2 = 2m_NE_r$ as
\begin{equation}
    \frac{d\sigma}{dE_\text{r}}=4\pi\alpha_B\alpha_\chi A^2\frac{2m_NE_\chi^2}{p_\chi^2\left(m_V^2+Q^2\right)^2}\lvert F(Q^2)\rvert^2
\end{equation}
where $p_\chi$ and $E_\chi$ are the momentum and energy of the incident dark-matter particle.  Timing of the $\pi$-DAR beam can differentiate the dark matter signal from CEvNS background as relativistic dark matter~\cite{Dutta:2020vop} produced by decay-in-flight of mesons in the SNS target would arrive coincident with the protons-on-target while the neutrino flux has a prompt $\nu_\mu$ and delayed $\nu_e/\bar{\nu}_\mu$ flux.  Production and scattering rates are predicted by the BdNMC event generator~\cite{deNiverville:2016rqh}.

\section{Analysis and results}

\begin{figure}[!bt]
\centering
\includegraphics[width=0.98\linewidth]{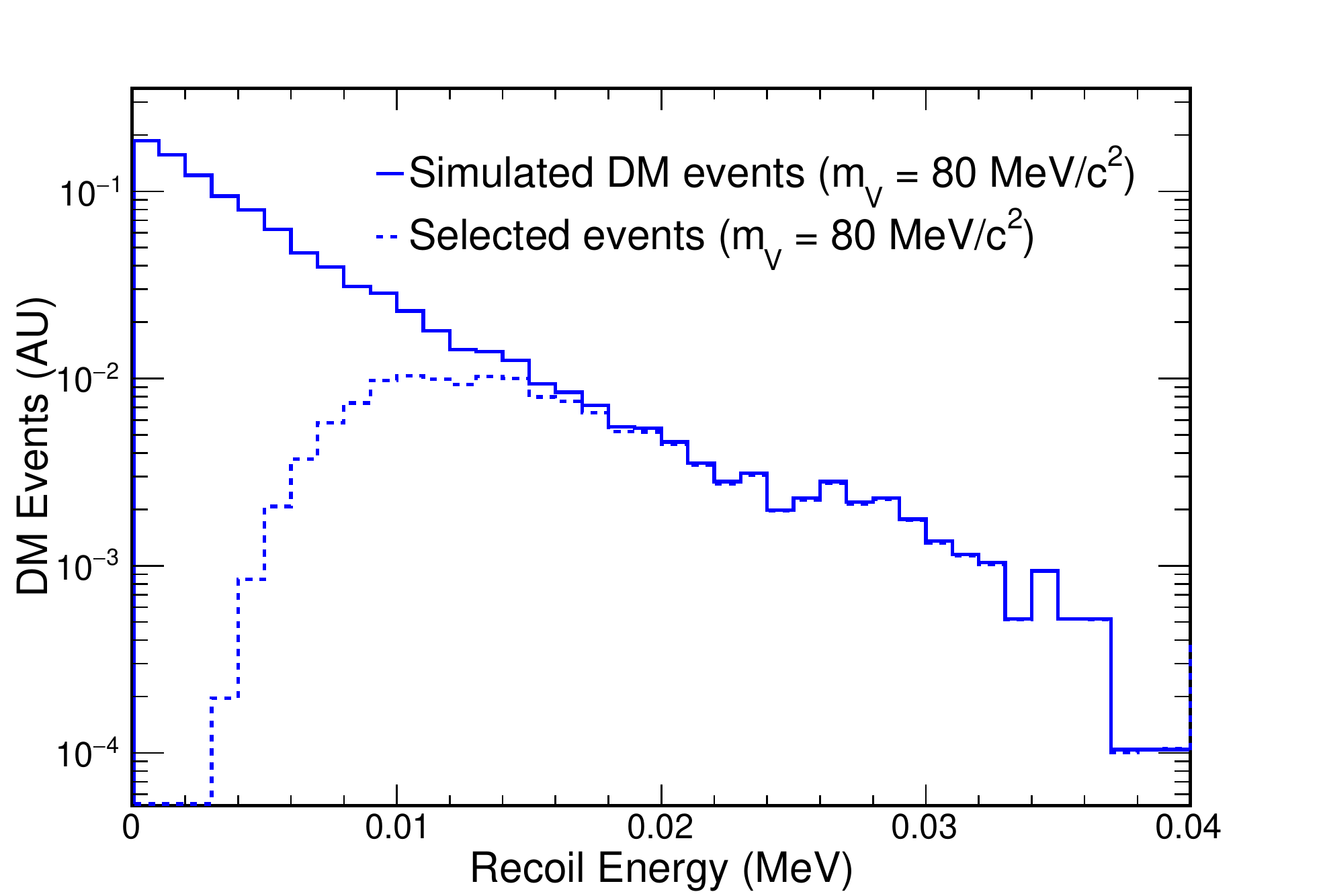}
\caption{Selection efficiency for simulated dark matter near the most sensitive mass.  With an approximately 9~keV$_\text{nr}$ threshold, the selection efficiency is 21$\%$ for dark matter.}
\label{fig:DMEff}
\end{figure}

The COHERENT CsI[Na] detector performance, along with discussion of event selection and background rates are presented here and discussed in detail in  \cite{Akimov:2021dab,COHERENT:2021pvd}.  Using calibration data from the 59.5-keV $\gamma$~peak from $^{241}$Am, we measured a light yield of 13.35~PE/keV$_\text{ee}$, or photoelectrons per keV of electronic recoil energy, in the CsI[Na] detector.  Nuclear recoil quenching was modeled by a polynomial fit to five neutron scattering measurements using a small CsI[Na] crystal with identical doping~\cite{COHERENT:2021pcd}.  A pulse-finding reconstruction was run on PMT waveforms to calculate a recoil time and energy for each event with 1~PE set to the average integral of SPE pulses.  For each beam spill, events were selected that have low background scintillation activity within the crystal prior to the beam spill and $\geq9$ reconstructed pulses.  These requirements resulted in a nuclear recoil threshold of $\approx9$~keV$_\text{nr}$ which was measured using $^{133}$Ba calibration data.  A four-parameter function was fit to the $^{133}$Ba data and also used to estimate the uncertainty in the threshold, $\approx 1$~keV$_\text{nr}$.  The efficiency is shown in Fig.~\ref{fig:DMEff} for simulated $m_V=80$~MeV/c$^2$ dark matter giving an average selection efficiency of 21$\%$ at this mass.  

Beam-unrelated backgrounds account for the majority, 78$\%$, of the total background below 60~PE and were measured in-situ with data out of time with the beam.  Due to afterglow scintillation activity within the crystal, the selection efficiency depends on recoil time as later recoils may be rejected due to spurious background activity earlier in the waveform.  The remaining backgrounds are neutron-related (2$\%$) whose normalizations were determined with data from a liquid scintillator housed in the CsI[Na] shielding collected before commissioning the CEvNS detector and CEvNS events (20$\%$).  The total systematic uncertainty on the CEvNS rate was 12$\%$ and was dominated by the neutrino flux uncertainty~\cite{COHERENT:2021yvp} with additional uncertainty calculated for quenching, the detection threshold model, form factor suppression, and background normalizations.  

\begin{figure}[!bt]
\centering
\includegraphics[width=0.98\linewidth]{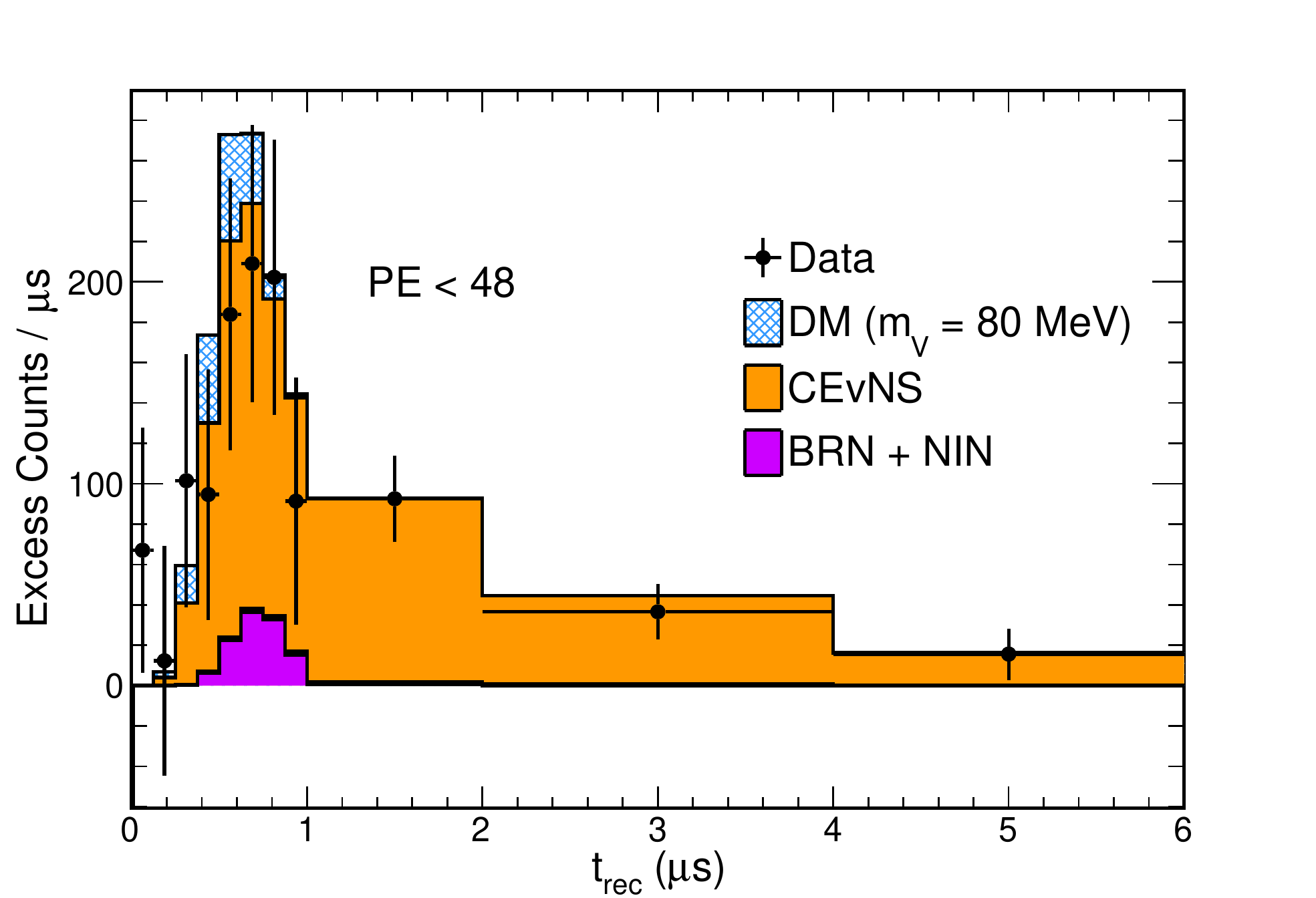}
\includegraphics[width=0.98\linewidth]{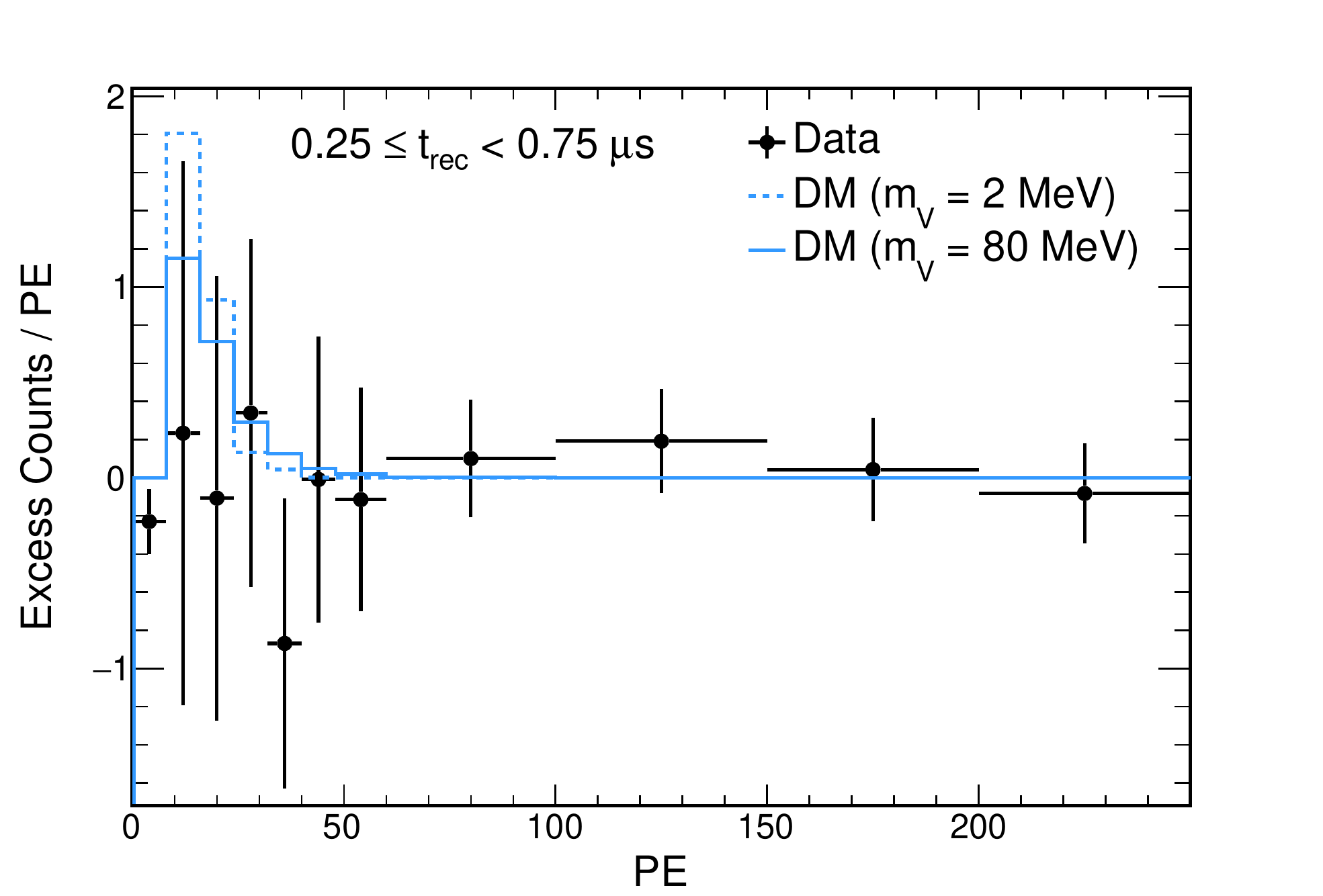}
\caption{The timing distribution of the beam-related excess (top) is shown with the 90$\%$ limit on dark matter content stacked with CEvNS and neutron (BRN and NIN) backgrounds at the best fit (with no dark-matter events).  The recoil energy distribution, with all backgrounds subtracted (including CEvNS), is also shown (bottom) with two assumptions of mediator mass.}
\label{fig:DataSpectra}
\end{figure}

The observed data were fit in both recoil energy and time for all selected events with $E_\text{rec}<250$~PE and $t_\text{rec}<6$~$\mu$s.  The observed data spectra are shown in Fig.~\ref{fig:DataSpectra}.  Though the dark-matter signal is expected to be in time with the arrival of the beam, a simultaneous fit to prompt and delayed data can constrain systematic uncertainties which are strongly correlated with CEvNS events in the prompt time window.  This affords a clear timing region of interest from 0.25 to 0.75~$\mu$s where the majority of dark matter is expected, with a background control region at recoil times above 0.75~$\mu$s.  This 2D fit ensures dark-matter searches at the SNS remain limited by statistical uncertainty even for the far-future sensitivity which will facilitate an increase in exposure by a factor of 1000.

\begin{figure}[!bt]
\centering
\includegraphics[width=0.98\linewidth]{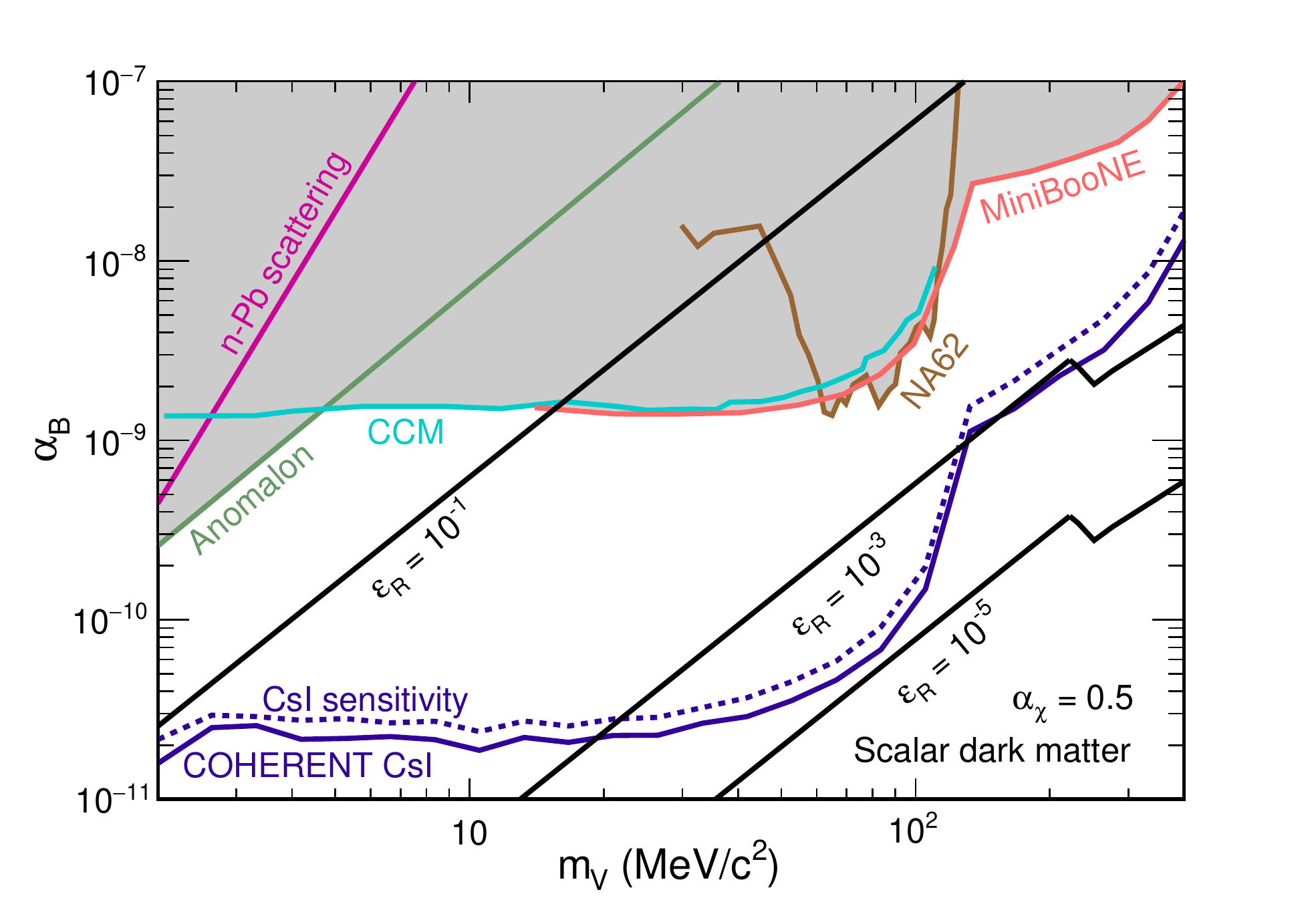}
\includegraphics[width=0.98\linewidth]{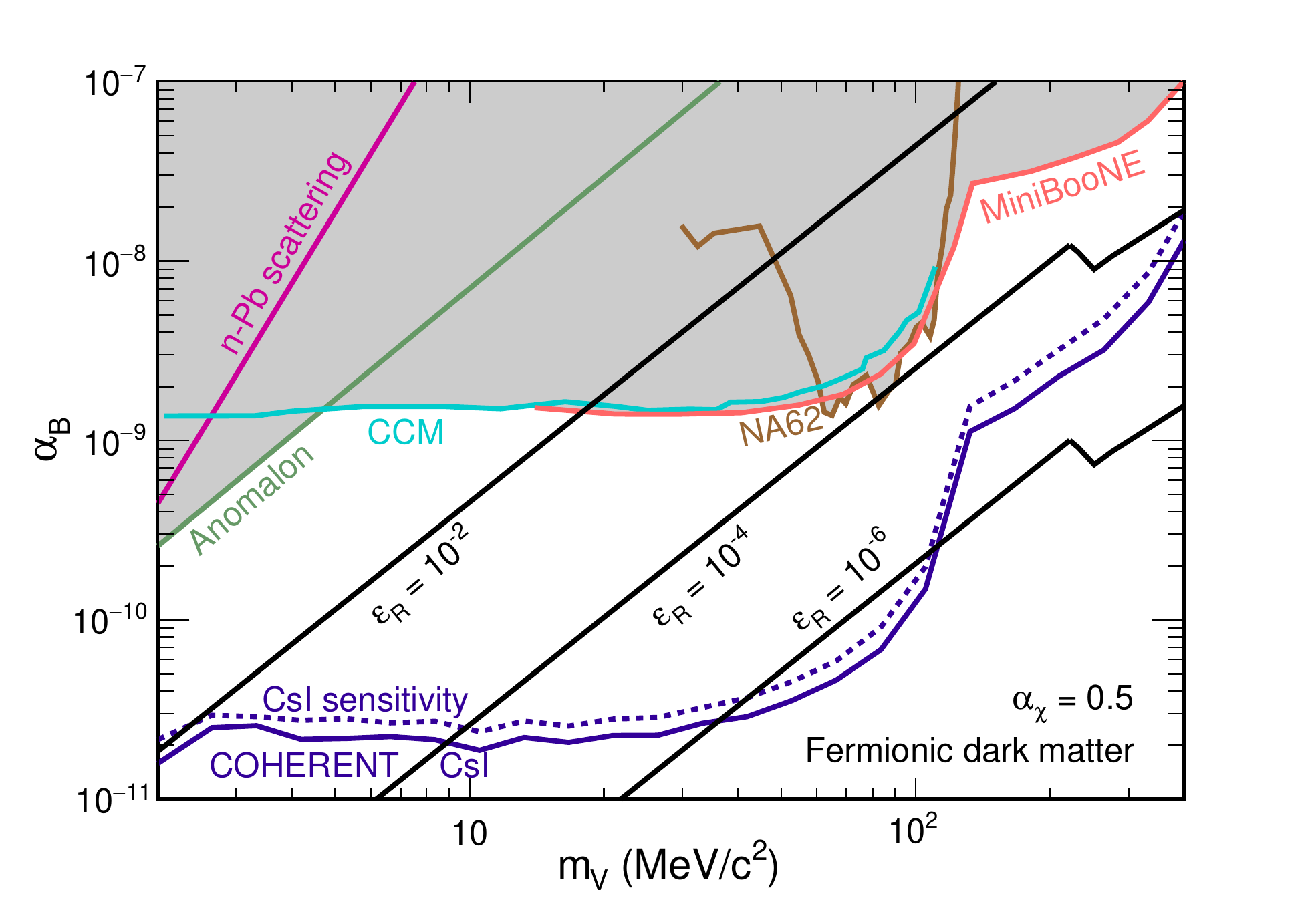}
\caption{Constraint on leptophobic parameter space from CsI[Na] data along with other experimental constraints for scalar (top) and fermionic (bottom) dark matter with $\alpha_\chi$ conservatively set to 0.5.  The couplings expected for the dark matter relic abundance are also plotted for different values of $\varepsilon_R$ in each case.}
\label{fig:DMresult}
\end{figure}

COHERENT data test leptophobic dark matter in the most conservative scenarios for $\varepsilon_R > 0$.  As such, couplings that match the relic abundance depend strongly on $m_V/m_\chi$.  For a fixed $m_V$, the expected dark-matter distribution does not change with varying $m_\chi$ for all $m_\chi<m_V/2$.  We thus show results in terms of $m_V$.  For a given value of $m_V$, $\alpha_\chi$ is assumed to be 0.5, as smaller values lead to stronger constraints and the model becomes non-perturbative at higher values.  A confidence interval is constructed for $\alpha_B$ using a log-likelihood spectral fit and the Feldman-Cousins method~\cite{Feldman:1997qc} at the 90$\%$ confidence level.  The resulting contour is shown in Fig.~\ref{fig:DMresult} compared to relic abundance targets with different assumptions of $\varepsilon_R$ for both scalar and fermionic dark matter.

The result places the strongest constraint on leptophobic dark matter over the entire mass range considered, $2<m_V<400$~MeV/c$^2$, improving the $\alpha_B$ bound by up to two orders of magnitude.  Throughout the entire region, the dark-matter relic abundance is excluded for $\varepsilon_R>0.01$ ($m_V/m_\chi>2.01$).  At the most sensitive mass, $m_V=$~80~MeV/c$^2$, the scalar result nearly reaches the $\varepsilon_R=10^{-5}$ line, the most conservative scenario for scalar leptophobic dark matter.  This region will be easily accessible with future COHERENT data.  This line is excluded for $\alpha_\chi<0.31$ for $m_V=80$~MeV/c$^2$.  Though there is no lower bound on the relic abundance in the fermionic case, the data can exclude the model for $\varepsilon_R=10^{-6}$ ($m_V/m_\chi-2\approx10^{-6})$ for $36<m_V<116$~MeV/c$^2$.  This constraint probes a dark matter flux 10000$\times$ lower than previously leading constraints, as many of the most sensitive probes of light dark matter are insensitive to leptophobic dark matter.  This result was the first performed by a detector sensitive to neutrino-induced CEvNS recoils in a $\pi$-DAR and was achieved with a small, 14.6~kg detector.  Future data from COHERENT will further probe leptophobic dark matter and can eliminate the model entirely in the scalar scenario with $m_V/m_\chi>2$.


\section{Conclusion}

COHERENT searched for leptophobic dark-matter particles at the SNS using the full dataset collected by the COHERENT CsI[Na] detector.  No dark matter signal was found, and strong constraints on the dark matter model are placed.  COHERENT places the most stringent limit to date for all mediator masses $2$~$<$~$m_V$~$<$~$400$~MeV/c$^2$.  For scalar dark matter mediated by a $80$~MeV/c$^2$ vector, this result nearly eliminates the studied dark-matter model for $m_V/m_\chi>2$.  Data from future COHERENT CEvNS detectors will strengthen constraints which can fully probe the scalar model and severely limit fermionic, leptophobic dark matter.

\section{Acknowledgements}

The COHERENT collaboration acknowledges the Kavli Institute at the University of Chicago for CsI[Na] detector contributions.  The COHERENT collaboration acknowledges the generous resources provided by the ORNL Spallation Neutron Source, a DOE Office of Science User Facility, and thanks Fermilab for the continuing loan of the CENNS-10 detector. We also acknowledge support from the Alfred~P. Sloan Foundation, the Consortium for Nonproliferation Enabling Capabilities, the National Science Foundation,  the Russian Foundation for Basic Research (proj.\# 17-02-01077 A), and the U.S. Department of Energy, Office of Science. Laboratory Directed Research and Development funds from ORNL and Lawrence Livermore National Laboratory also supported this project.  This research used the Oak Ridge Leadership Computing Facility, which is a DOE Office of Science User Facility.  The work was supported by the Ministry of Science and Higher Education of the Russian Federation, Project Fundamental properties of elementary particles and cosmology No. 0723-2020-0041.

\end{document}